\documentclass[aps,prl,twocolumn,showpacs,superscriptaddress]{revtex4}
\usepackage{graphicx,color,epsfig}
\usepackage{amssymb,amsmath}
\begin{document}

\title{A model for cross-cultural reciprocal interactions through mass media}
\author{J. C. Gonz\'alez-Avella}
\affiliation{Instituto de F\'isica, Universidade Federal do Rio Grande do Sul, 91501-970 Porto Alegre, Brazil}
\author{M. G. Cosenza}
\affiliation{Centro de F\'isica Fundamental, Universidad de los Andes, M\'erida, Venezuela.}
\author{M. San Miguel}
\affiliation{IFISC, Instituto de F\'isica Interdisciplinar y Sistemas Complejos (CSIC-UIB), E-07122 Palma de Mallorca, Spain} \date{PLoS ONE \textbf{7} (12): e51035. doi:10.1371/journal.pone.0051035 (2012)}

\begin{abstract}
We investigate the problem of cross-cultural interactions through mass media in
a model where two populations of social agents, each with its own internal dynamics, get information 
about each other through reciprocal global interactions.
As the agent dynamics, we employ Axelrod's model for social
influence. The global interaction fields correspond
to the statistical mode of the states of the agents and represent mass media messages on the cultural trend
originating in each population. Several phases are found in the collective behavior of 
either population depending on parameter values: two homogeneous phases, one having the state 
of the global field acting on that population, and the other consisting of a state different from that reached by the applied global field; and a disordered phase. In addition, the system displays nontrivial effects: (i) the emergence of a largest minority group of appreciable size sharing a state different from that of the applied global field; (ii) the appearance of localized ordered states for some values of parameters 
when the entire system is observed, consisting of one population in a homogeneous state and the other in a disordered state. 
This last situation can be considered as a social analogue to a chimera state
arising in globally coupled populations of oscillators.
\end{abstract}

\maketitle

\section{Introduction}
The study of cross-cultural experiences through mass-mediated contact is a topic of much interest in the Social Sciences \cite{Elamar,Bryant,Nova,Yaple,Chinos}. Many of those studies have focused on the effects of cultural product consumption on audience beliefs, emotions, and attitudes toward the group originating these cultural products. For instance, several works have investigated the process by which international audiences develop American values, norms and stereotypes about America through the experience of watching American television series \cite{Tan,Weiman,Zaha}. 
Other works have explored the political impact of international television across borders \cite{Hunter}. 
The expansion of broadcasting and telecommunication industries in recent times has led to an increase in the 
exchange of mass media products across countries and social groups. As a consequence, people of different groups that may have had little direct contact with each other can, however, have access to their reciprocal mass media messages.
For example, the growth of media channels in East Asia has brought changing patterns of cultural consumption: younger generations in China are drawn to Korean pop stars; Korean people have begun to collect Chinese films; Japanese audiences await the broadcast of non-Japanese Asian dramas \cite{Chinos}.

In the current research in complex systems, there is also much  
interest in the investigation of models of social dynamics \cite{Castellano}.  
Many of these systems have provided scenarios for investigating new forms of
interactions and for studying new collective phenomena 
in non-equilibrium systems \cite{Marro,Weid,Sucheki,Stauffer,Vilone1,Zimmerman,Maxi,Holme,Centola,Barto}. 
In this context, the model introduced by Axelrod \cite{Axel} 
to investigate the dissemination of culture among interacting agents in a society 
has attracted much attention from physicists \cite{CMV,Vilone,Klemm,Klemm2,JC1,JC,Klemm3,Kuperman,Candia,Galla,Gracia,Zhang}.
In this model, the agent-agent interaction rule is such that no interaction exists
for some relative values characterizing the states of the agents that compose the system. This
type of interaction is common in social and biological systems where there is often some bound or restriction
for the occurrence of interaction between agents, such as a similarity condition for the state
variable \cite{Mikhailov,Deff,Amblard,Krause,Zanette}. 

In particular, the effects of local and global mass media on a social group have been studied by using Axelrod's model \cite{Shibanai,JC1,JC,NJP}. Some different formalisms for mass media based on Axelrod's model have also been proposed \cite{Candia2,Arezky1,Arezky2}. 

In this paper we investigate the problem of cross-cultural interactions through mass media in
a model where two separated social groups, each with its own internal dynamics, get information about each other solely through reciprocal global interactions.
We address the question of whether two societies subject to reciprocal mass media interactions become more similar to each other or if they can mantain some diversity. 
Specifically, our system consists of 
two populations of social agents whose dynamics 
is described by Axelrod's model, mutually coupled through global interactions.
The global interactions act as fields that can be interpreted as 
mass media \cite{JC,Gar}. In our model, the mass media content reaching one population corresponds to the statistical mode or cultural trend  originated in the other population, and viceversa. 

The existence of non-interacting states in the dynamics, as well as
the competition between the time scales of local agent-agent interactions
and  the responses of the endogenous global fields,
lead to nontrivial collective behaviors, such as the emergence of a largest minority group in a population, 
sharing a state different from that of the applied global field,
and the occurrence of localized ordered states. 
In this last case, one population reaches a homogeneous state while 
several states coexist on the other. This situation can be considered as a social analogue to a chimera state
arising in globally coupled populations of oscillators \cite{Kuramoto,Strogatz,Abrams,Laing,Scholl,Showalter}.

In Sec.~1 we present the model for two interacting populations of social agents and characterize the collective behavior on the space of parameters of the system. 
The nature of the observed localized ordered states is investigated in Sec.~2. Section~3 contains the conclusions of this work.

\section{The model}
We consider a system of $N$ agents consisting of two populations or subsets: $\alpha$ and $\beta$,  with sizes $N_\alpha$ and $N_\beta$, such that $N=N_\alpha+N_\beta$. The fraction of agents in subset $\alpha$ is $N_\alpha/N$ and that in subset $\beta$ is $N_\beta/N$.

Each subset consists of a fully connected network, i. e., every agent can interact with any other within a subset.
We employ the notation $[z]$ to indicate ``or $z$''.
The state of agent $i \in \alpha [\beta]$ is given by an $F$-component vector $x_{\alpha[\beta]}^f(i)$,
$(f = 1, 2, \ldots,F)$, where each component can take any of $q$ different values 
$x_{\alpha[\beta]}^f(i) \in \{0, 1, . . . , q-1\}$. 

Let us denote by $M_\alpha=(M_\alpha^1,\ldots,M_\alpha^f,\ldots,M_\alpha^F)$ and $M_\beta=(M_\beta^1,\ldots,M_\beta^f,\ldots,M_\beta^F)$ 
the global fields  defined as the statistical modes of the states in the subsets $\alpha$ and  $\beta$, respectively, at a given time. This means that the component $M_{\alpha[\beta]}^f$ 
is assigned the most abundant value exhibited by the $f$th component of all the state vectors $x_{\alpha[\beta]}^f(i)$ in the subset $\alpha[\beta]$. If the maximally abundant value is not unique, one of the possibilities
is chosen at random with equal probability. In the context of social dynamics, these global fields can be interpreted 
as mass media messages about ``trends'' originated in each population.

\begin{figure}[h]
\begin{center}
\includegraphics[width=0.8\linewidth,angle=0]{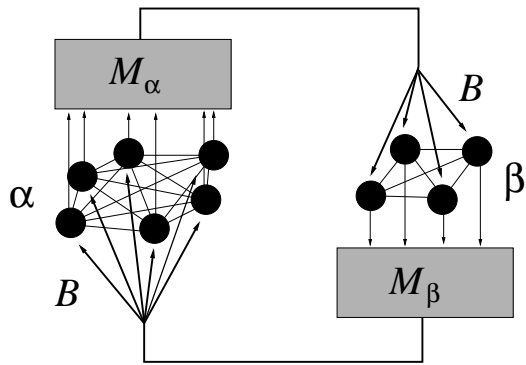}
\end{center}
\caption{Representation of two populations $\alpha$ and  $\beta$ interacting through their reciprocal global fields
$M_\alpha$ and $M_\beta$, each acting with intensity $B$.}
\label{F1}
\end{figure}

Each agent in subset $\alpha$ is subject to the influence of the global field $M_\beta$, 
and each agent in subset $\beta$ is subject to the influence of the global field $M_\alpha$. 
Figure~\ref{F1} shows the configuration of the two populations subject to the influence of their 
reciprocal global fields.

Starting from random initial conditions in each subset, at any given time, a randomly selected agent in subset $\alpha[\beta]$ 
can interact either with the global field  $M_{\beta[\alpha]}$ or with any other agent belonging to 
$\alpha[\beta]$. The interaction in each case takes place according to the dynamics of Axelrod's cultural model.

The dynamics of the system is defined by iterating the following steps:
\begin{enumerate}
\item Select at random an agent $i\in \alpha$ and a agent $j \in \beta$.
\item Select the source of interaction:
with probability $B$, agent $i \in \alpha$ interacts with field $M_\beta$ and agent $j \in \beta$ interacts with field $M_\alpha$, while
with probability $1-B$,  $i$ interacts with $k \in \alpha$ selected at random and $j$ interacts with $l \in \beta$ also selected at random. 
\item Calculate the overlap (number of shared components) between agent $i \in \alpha$ and its source of interaction, given by $d_\alpha= \sum_{f=1}^F \delta_{x_\alpha^f(i),y^f}$, where $y^f=M_\beta^f$ if the source is the field $M_\beta$, or $y^f=x_\alpha^f(k)$ if the source is agent $k \in \alpha$. Similarly, calculate the overlap $d_\beta= \sum_{f=1}^F \delta_{x_\beta^f(j),y^f}$, where $y^f=M_\alpha^f$ if the source is the field $M_\alpha$, or $y^f=x_\beta^f(l)$ if the source is agent $l\in \beta$. Here we employ the delta Kronecker function, $\delta_{x,y}=1$, if $x=y$;  $\delta_{x,y}=0$, if $x\neq y$.
\item If $0 < d_\alpha <F$, with probability $\frac{d_\alpha}{F}$ choose $g$ such that $x_\alpha^g(i)\neq y^g$ and set 
$x_\alpha^g(i)=y^g$; if $d_\alpha=0$ or $d_\alpha=F$, the state $x_\alpha^f(i)$ does not change. 
If $0 < d_\beta <F$, with probability $\frac{d_\beta}{F}$ choose $h$ such that $x_\beta^h(j)\neq y^h$ 
and set $x_\beta^h(j)=y^h$; if $d_\beta=0$  or  $d_\beta=F$, the state $x_\beta^f(j)$ does not change.
\item If the source of interaction is $M_{\beta[\alpha]}$, update the fields $M_\alpha$ and $M_\beta$.
\end{enumerate}

The strength of each field $M_\alpha$ and $M_\beta$ is represented by the parameter $B \in [0, 1]$ that measures the
probability for the agent-field interactions. 
Step $5$ characterizes the time scale for the updating of the global fields in our model.
In general, agents in one population do not have instantaneous knowledge of the state of the global field of the other population, but only when they effectively interact with that global field.
The non-instantaneous updating of the global fields expresses the delay with which 
a population acquires
knowledge about the other through
the only available communication channel between them,
as described in many cross-cultural interactions through mass media \cite{Chinos}. 
In our case, as the value of the parameter $B$ increases, both the intensity of the global fields 
and the updating rate of their states increase.

Under the mutual coupling, both populations, $\alpha$ and $\beta$ form domains of different sizes in the asymptotic state.

\begin{figure}[h]
\begin{center}
\includegraphics[width=0.65\linewidth,angle=0]{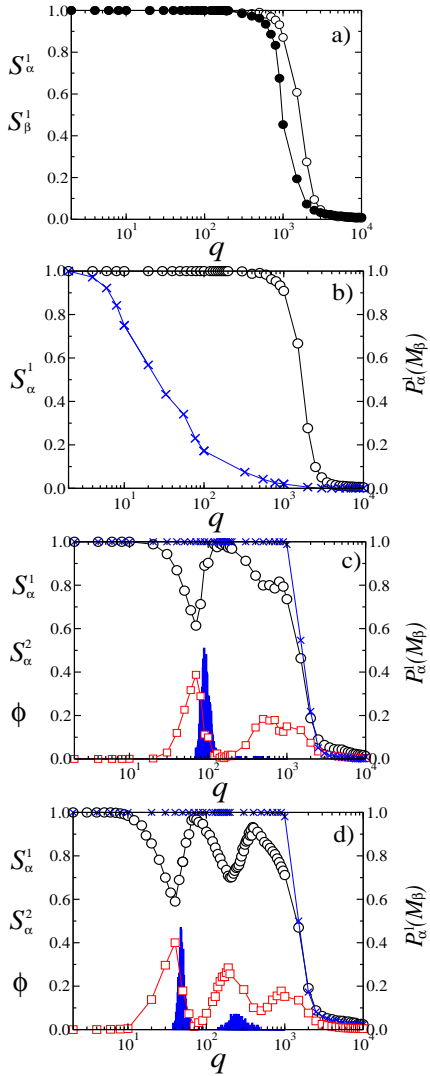}
\end{center}
\caption{$S_\alpha^1$, $S_\alpha^2$, and $P_\alpha^1(M_\beta)$ as functions of $q$, with $F=10$, for different values of $B$. System size is $N=800$ with partition $N_\alpha=0.6N$. Each data point is the result of averaging over $100$ random realizations of initial conditions. (a) $S_\alpha^1$ (open circles), $S_\beta^1$ (solid circles); with $B=0$. (b) Left vertical axis: $S_\alpha^1$ (open circles); right vertical axis: $P_\alpha^1(M_\beta)$ (crosses); fixed $B=0.001$. Phases I and II. (c) Left vertical axis: $S_\alpha^1$ (open circles), $S_\alpha^2$ (open squares); right vertical axis: $P_\alpha^1(M_\beta)$ (crosses); fixed $B=0.05$. Phases I and IV. (d) Left vertical axis: $S_\alpha^1$ (open circles), $S_\alpha^2$ (open squares);  right vertical axis: $P_\alpha^1(M_\beta)$ (crosses); fixed $B=0.25$. Phase III occurs for values $q>q_c=2500$, independent of $B$.
The bars in (c) and (d) indicate the probability $\phi$ of finding a localized ordered state in the system
as a function of $q$ for the given value of the intensity $B$.}
\label{F2}
\end{figure}

A domain is a set of connected agents that share the same state.
A homogeneous or ordered phase in a population corresponds to $d(i,j)=F$, $\forall i, j$. There are $q^F$ equivalent
configurations for this ordered phase. In an inhomogeneous or disordered phase in a population several domains coexist.
The sizes of these domains within each population are ranked by the index $r$: $r=1$ corresponding to the largest domain, 
$r=2$ indicates the second largest domain, etc.
To characterize the collective behavior of the system, we define the following macroscopic quantities:
(i) the average normalized size (divided by $N_{\alpha[\beta]}$) of the domain in $\alpha[\beta]$ whose size has rank $r$, denoted by $S_{\alpha[\beta]}^r$; 
(ii) the probability that the largest domain in $\alpha[\beta]$ has a state equal to $M_{\beta[\alpha]}$, designed by $P_{\alpha[\beta]}^1(M_{\beta[\alpha]})$.

\begin{figure}[h]
\begin{center}
\includegraphics[width=0.95\linewidth,angle=0]{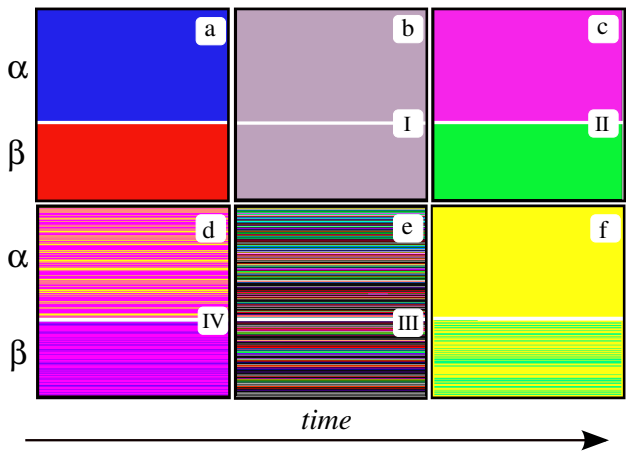}
\end{center}
\caption{Each panel displays an 
asymptotic state (vertical axis) of the agents in the interacting populations $\alpha$ (upper part) 
and $\beta$ (lower part) 
vs. time (horizontal axis), corresponding to a different phase in the system. 
Each value of the state variable of an agent is represented by a different color. 
Population sizes are $N_\alpha=0.6N$, $N_\beta=0.4N$, with $N=800$. 
(a) $B=0, q=80$ (no coupling). 
(b) $B=0.001, q=80$ (phase I). (c) $B=0.001, q=100$ (phase II). (d)$B=0.05, q=80$ (phase IV). 
(e) $B=0.25, q=2500$ (phase III). (f) $B=0.05, q=80$ (localized ordered state).}
\label{F3}
\end{figure}

Figure~\ref{F2} shows various of these quantities as functions of the parameter $q$, for different values of $B$.
In this paper we fix the parameter value $F=10$.
In the absence of global fields (Fig.~\ref{F2}(a)),
i.e. $B = 0$,  we have two uncoupled and independent subsets; 
each subset spontaneously reaches an ordered phase, characterized by $S_\alpha^1=1$ and $S_\beta^1=1$, 
for values $q < q_c$, and a disordered phase, corresponding to $S_\alpha^1 \simeq  0$ and $S_\beta^1 \simeq  0$, 
for $q > q_c$, where $q_c$ is a critical point that depends on the subset size in each
case, $q_c (\alpha)\sim N_\alpha$ \cite{Fede}. Figure~\ref{F3}(a) shows the asymptotic pattern in this case.

For $B\rightarrow 0$ and $q < q_c$, each population reaches an ordered state with $S_\alpha^1=1$, 
as shown in Fig.~\ref{F2}(b).
However, in this situation the spontaneous order emerging in subset $\alpha$ for parameter
values $q < q_c$ due to the agent-agent interactions competes with the order being imposed
by the applied global field $M_\beta$. 
For some realizations of initial conditions, the global field $M_\beta$ imposes its state on subset $\alpha$ 
and,
correspondingly, the field  $M_\alpha$ imposes its state on subset $\beta$. As a consequence, both subsets 
reach the same state with  $M_\alpha=M_\beta$. An asymptotic state corresponding to this situation 
is displayed
in Fig.~\ref{F3}(b). We refer to this state as phase I. However, the ordered state in subset $\alpha[\beta]$ does not always 
correspond to the state of the global field $M_{\beta[\alpha]}$ being applied to $\alpha[\beta]$. This is
revealed by the probability $P_\alpha^1(M_\beta)$ shown in Fig.~\ref{F2}(b) that
measures the fraction of realizations that 
the largest domain in $\alpha$ has a state equal to $M_\beta$. We find $P_\alpha^1(M_\beta)<1$ for a range of values $q<q_c$.
Thus, in this case there is a probability that 
subsets $\alpha$ and $\beta$ can reach ordered states different from each other,
i. e., $M_\beta \neq M_\alpha$. Figure~\ref{F3}(c) illustrates the asymptotic states 
in this case. We denote this situation as phase II. 

Figures~\ref{F2}(c) and \ref{F2}(d) show both $S_\alpha^1$ and $S_\alpha^2$ as functions of $q$ for greater values of $B$. 
The quantity $S_\alpha^1$ in Fig.~\ref{F2}(c) displays a local minimum at some value of $q$ that depends on $B$. 
This local minimum of $S_\alpha^1$ is associated to a local maximum value of $S_\alpha^2$, such that $S_\alpha^1+S_\alpha^2 \approx 1$ for $q < q_c$. Therefore, two majority domains form in subset $\alpha$ for $q < q_c$.  Fig.~\ref{F2}(c) also shows that the probability $P_\alpha^1(M_\beta)=1$,
indicating that the state of the largest group in $\alpha$ is always equal to that imposed by the field $M_\beta$. But the  second largest group that occupies almost the rest of subset $\alpha$ reaches a state different from $M_\beta$. Thus, the value of $q<q_c$ for which $S_\alpha^1$ has a local minimum is related to the emergence of a second largest domain ordered against the global field $M_\beta$. The corresponding asymptotic pattern is shown in Fig.~\ref{F3}(d). We call this configuration phase IV.
Figure~\ref{F2}(d) reveals that, for larger values of $B$, various local minima of $S_\alpha^1$ can occur at some values of $q$. This local minima of $S_\alpha^1$ correspond to local maxima of $S_\alpha^2$ and to the emergence of a second largest domain in $\alpha$ ordered against the field $M_\beta$. 
The raise of a largest minority group at some values of $q$ is a manifestation of the tendency towards the spontaneous order related to the agent-agent interactions. 
For values $q > q_c$, both populations reach disordered states $\forall B$, characterized by 
$S_\alpha^1 \simeq  S_\beta^1 \simeq  0$. The disordered behavior of the system is denoted by phase III 
and the corresponding pattern is displayed in Fig.~\ref{F3}(e).

\begin{figure}[h]
\begin{center}
\includegraphics[width=0.9\linewidth,angle=0]{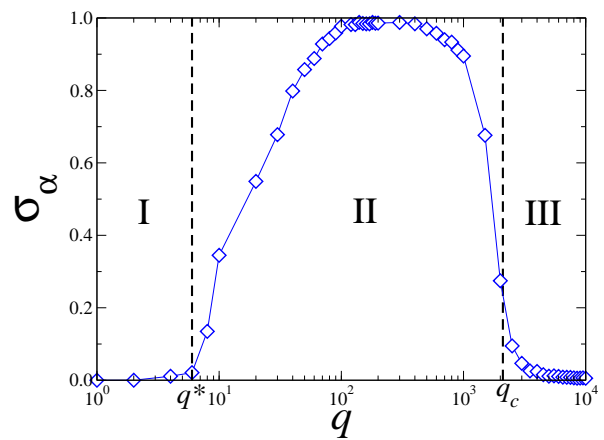}
\end{center}
\caption{The quantity $\sigma_\alpha=(1-P_\alpha^1(M_\beta)) S_\alpha^1$ as a function of $q$ for a fixed value 
$B=0.0005$, with $F=10$. The critical values $q^*$ and $q_c$, as well as the regions where phases I, II, and III occur, are indicated. 
System size is $N=800$ with partition $N_\alpha=0.6N$. Each data point is averaged over $100$ realizations of initial conditions.}
\label{F5}
\end{figure}

To characterize phase II, we plot in Fig.~\ref{F5} the quantity $\sigma_\alpha=(1-P_\alpha^1(M_\beta)) S_\alpha^1$
as a function of $q$, for a fixed value $B=0.0005$. For $q < q^*\approx 10$, the state of the largest domain in $\alpha$ corresponds to the state of the field $M_\beta$, i.e. $P_\alpha^1(M_\beta)=1$ and $S_\alpha^1=1$, indicating the presence of phase I, and thus
$\sigma_\alpha=0$. For $q^*<q<q_c$, the largest domain in $\alpha$ no longer possesses the state of the field $M_\beta$
but another state non-interacting with this field, i.e. $P_\alpha^1(M_\beta)<1$ and $S_\alpha^1=1$, and therefore $\sigma_\alpha>0$, characterizing phase II. For $q>q_c$, $S_\alpha^1 \rightarrow 0$ and
$\sigma_\alpha=0$.

We note that phase II occurs for small values of $B$, where the time scale for the agent-agent interaction dynamics is smaller than the corresponding time scale for the agent-field dynamics. This means that
the state of the global field 
does not vary much in comparison to the changes taking place in the states of the agents and, therefore, the global field behaves approximately as a fixed external field with little influence on the system. As a consequence the system can spontaneously order in a state different from that of the global field if $q<q_c$ is sufficiently large, giving rise to phase II. For increasing values of $B$, the updating of the global fields and the agent-agent dynamics have comparable time scales and, therefore, the state of the fields corresponds to that of the largest domain in each subset, yielding regions of both phase I and phase IV. 

\begin{figure}[h]
\begin{center}
\includegraphics[width=0.9\linewidth,angle=0]{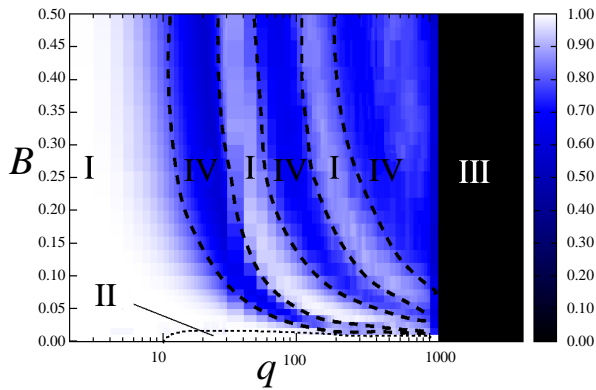}
\end{center}
\caption{Phase diagram of population $\alpha$ on the space of parameters $(B,q)$, with $F=10$. System size is $N=800$ 
with partition $N_\alpha=0.6N$. Each data point is averaged over $100$ realizations of initial conditions. 
The color code represents the value of the normalized largest domain size $S_\alpha^1$, from black ($S_\alpha^1=0$) 
to white ($S_\alpha^1=1$). The regions where
the different phases occur are labeled and separated by slashed lines: phase I (both populations share same homogeneous
state); 
phase IV (partially ordered, emergence of second group); phase III (disordered), and phase II (each population in a different homogeneous
state).
Localized ordered states can occur in the transitions from phase IV to phase I.}
\label{F4}
\end{figure}

The collective behavior of either of the two subsets coupled through their reciprocal global fields can be characterized by four phases on the space of parameters $(B,q)$, as shown in Fig.~\ref{F4} for subset $\alpha$: (I) a homogeneous, ordered  phase, for which $S_\alpha^1 \sim 1$ and $P_\alpha^1(M_\beta)=1$;
(II) an ordered phase in a state orthogonal to the applied global field, such that  $S_\alpha^1 \sim 1$ and 
$P_\alpha^1(M_\beta)<1$; (III) a disordered phase for $q>q_c$, for which $S_\alpha^1 \simeq  0$; and
(IV) a partially ordered phase, where $S_\alpha^2>0$ and $S_\alpha^1+S_\alpha^2 \approx 1$,
$P_\alpha^1(M_\beta)=1$, characterized by the emergence of a second largest domain ordered in a state different from field $M_\beta$. 

The phase diagram of Fig.~\ref{F4} reveals that the interaction through reciprocal, evolving global fields can lead to nontrivial effects in certain cases. 
For example, for a fixed value  $q=20$,
the global field can impose its state to the system (phase I) only for
a range of intermediate values of the intensity $B$.

\begin{figure}[h]
\begin{center}
\includegraphics[width=0.74\linewidth,angle=0]{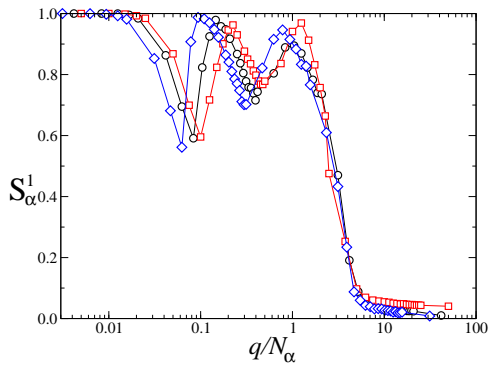}
\end{center}
\caption{Normalized size of largest domain $S_\alpha^1$ as a function of $q/N_\alpha$ with fixed intensity $B = 0.1$, for different population sizes: $N_\alpha=0.5N$ (squares);  $N_\alpha=0.56N$ (circles);  $N_\alpha=0.8N$ (diamonds).  System size is $N = 800$ and $F=10$.}
\label{F8}
\end{figure}

We have checked the behavior of the system for different population sizes $N_\alpha$ and $N_\beta$. Figure~\ref{F8} shows the quantity $S_\alpha^1$ as a function of $q/N_\alpha$ with fixed coupling $B$, for different values of $N_\alpha$. We see that the critical point for the transition to phase III scales as $q_c \sim N_\alpha$, as expected  \cite{Fede}, and that the qualitative collective behavior represented in the phase diagram of Fig.~\ref{F4} is independent of the sizes of the partitions into two populations. Since $N_\alpha \propto N$, the collective behavior of the system is also independent of the size $N$, and $q_c \sim N$, according to Fig.~\ref{F8}.

\section{Localized ordered states}
In addition to phases I and II that display homogeneous states for both subsets $\alpha$ and $\beta$, there are configurations where homogeneous states can take place in only one subset, while the other is inhomogeneous, for some values of parameters. We refer to this configuration as \textit{localized ordered states}. These states are characterized by
$S_\alpha^1 \; [S_\beta^1]=1$ and $S_\beta^1 \; [S_\alpha^1]=u<1$. Figure~\ref{F3}(f) displays the asymptotic  state of the system in this case.
In contrast to the four phases that can be characterized in a subset, the ordered collective states can only be defined
by considering both subsets simultaneously, i.e., it requires the observation of the entire system.

To elucidate the nature of these states, we calculate the probability $\phi$ of finding a localized ordered state in the  system as a function of $q$ in Figs. 2(c) and 2(d), employing the criterion $u \leq 0.6$. In both figures, 
there are ranges of the parameter $q$ where localized ordered states can occur;  
the probability $\phi$ 
is maximum near the values of $q$ that correspond to local minima of $S_\alpha^1$ (and local maximum values of $S_\alpha^2$).

\begin{figure}[h]
\begin{center}
\includegraphics[width=0.95\linewidth,angle=0]{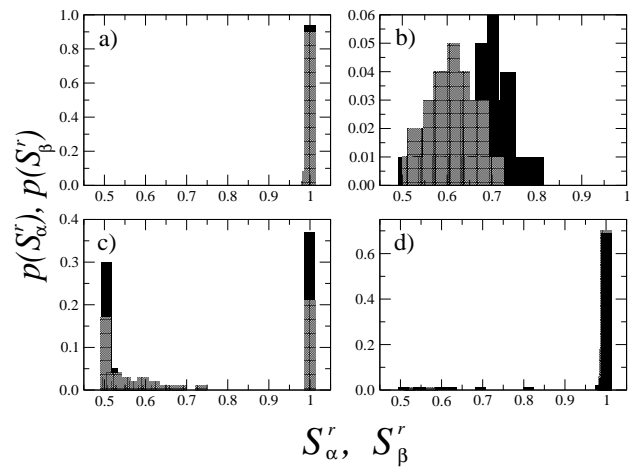}
\end{center}
\caption{Probability distributions $p(S_\alpha^r)$ and $p(S_\beta^r)$, $\forall r$, of normalized domain sizes for both  populations $\alpha$ (black bars) and $\beta$ (grey bars), calculated over 
$100$ realizations of initial conditions,
with fixed intensity $B=0.05$, $F=10$, and for different values of the number of options $q$. (a) $q=10$ (phase I); (b) $q=70$ (phase IV); (c) $q=90$ (localized ordered states); (d) $q=115$ (phase I).}
\label{F7}
\end{figure}
 
Figure~\ref{F7} shows the probability distributions $p(S_\alpha^r)$ and $p(S_\beta^r)$, $\forall r$, of the normalized domain sizes for both subsets $\alpha$ and $\beta$, calculated over $100$ realizations of initial conditions, for different values of $q$, and 
with fixed $B=0.05$ corresponding to Fig.~2(c). 
Figure~\ref{F7}(a) exhibits the probabilities $p(S_\alpha^r)$ and $p(S_\beta^r)$ when either subset is in phase I with $q=10$, characterized by the presence of
one large domain whose size is of the order of the system size $S_\alpha^1 [S_\beta^1] \sim 1$, in agreement with Fig.~3(b). Figure~\ref{F7}(b) shows $p(S_\alpha^r)$ and $p(S_\beta^r)$ associated to phase IV ($q=70$), where the size of the largest domain in either subset never reaches the system size due to the appearance of a second group, as displayed in Figs.~2(c) and 3(d).  Figure~\ref{F7}(c) shows the probabilities $p(S_\alpha^r)$ and $p(S_\beta^r)$ for $q=90$. In this case either subset can reach an ordered configuration, $S_\alpha^1 [S_\beta^1] \sim 1$,
or an inhomogeneous state $(S_\alpha^1 [S_\beta^1] < u)$. This corresponds to the appearance of localized ordered states in the system. For $q=115$, we find again a probability distribution typical of phase I.

The localized ordered states are analogous to chimera states observed in two populations of dynamical oscillators
having global or long range interactions, where one population in a coherent state coexist with the other in a incoherent state \cite{Kuramoto,Strogatz,Abrams,Laing,Scholl,Showalter}. In a chimera state, one part of a spatially extended system presents a coherent or synchronized  behavior while another part is desynchronized. 

Note that the regions of parameters where localized ordered states can emerge in our system lie 
between phase IV and phase I states. In fact, the configuration of localized ordered states shares features of both phase I and phase IV; they can be considered as transition configurations between phase IV and phase I states. 

\section{Discussion}
We have investigated the collective behavior of a 
system consisting of 
two populations of social agents, 
mutually coupled through global fields,
as a model for cross-cultural interactions via mass media. 
Specifically, we have employed Axelrod's model for social influence as the interaction dynamics. 

The global interaction field associated to each population corresponds to the statistical mode of the states of the agents.
In the context of social dynamics, this global autonomous field can be interpreted as mass media messages about ``trends'' or stereotypes originated in one population that are transmitted to the other population. Thus, our system can represent
cross cultural interactions between two separated social groups, each with its own internal dynamics, but getting information about each other solely through their mass media messages \cite{Chinos}. 

We have found several phases on either subset depending on parameter values: two homogeneous phases, one having the state 
of the global field acting on that subset (phase I), and the other consisting of a state different from that reached by the applied global field (phase II); a partially ordered phase characterized by the emergence of a second largest domain ordered in a state different from the global field (phase IV); and a disordered phase (III). 

States similar to phases I, II, and III are also observed for some regions of parameters in a system of social agents subject to an external fixed field \cite{NJP}.  In the present model with non-instantaneous updating of the fields, for small values of $B$, the global evolving field varies very slowly in comparison to the changes in the states of the agents in a subset due to their mutual interactions. In this case, the global evolving field behaves as a fixed external field acting on the population.

However, for larger values of $B$, the adaptive nature of the global fields induce two new phenomena in some range of $q<q_c$ on each population. One is the emergence of a largest minority group of appreciable size having a state different from that of the applied field (phase IV). The other corresponds to the appearance of localized ordered states
when the entire system is observed, consisting of one population in a homogeneous state and the other in an disordered state. These configurations occur with a probability
that depend on both $B$ and $q$ and appear as transitions states from phase IV to phase I. 
These localized ordered states are analogous to the chimera states that have been found in networks of coupled oscillators
having global interactions, where a subset of the system reaches a coherent state while another subset remains incoherent \cite{Abrams,Laing}. The recent experimental discovery of such chimera states has fundamental implications as it shows that localized order and structured patterns can emerge from otherwise structureless system \cite{Showalter,Roy}. 
As noted in Ref.~\cite{Abrams}, analogous symmetry breaking is observed in dolphins and other animals
that have evolved to sleep with only half of their brain at a time: neurons exhibit
synchronized activity in the sleeping hemisphere and desynchronized activity in the
hemisphere that is awake \cite{Lima}. 

From a social perspective, our model shows that cross cultural reciprocal interactions through mass media 
do not always lead to the imposition over one population 
of the cultural trends being transmitted by the media of another population.  
A group possessing a cultural state different from that of the mass media message can spontaneously
emerge in the first population. Under some circumstances, such group can encompass the entire population
(phase II), or it can constitute the largest minority in that population (phase IV). 

The behaviors reported here should also be expected in other non-equilibrium systems possessing 
non-interacting states, such as social and biological systems whose dynamics usually possess a bound
condition for interaction \cite{Deff}. This includes models of motile elements
in population dynamics, such as swarms, fish schools, bird flocks and bacteria colonies \cite{Mikhailov,Huth,Vicsek,Chate,Kaneko,Kenkre}.
Future extensions of this work involves the consideration of complex network structures within each population and 
the investigation of communities, where the interaction between populations occurs through a few elements rather than
a global field.

\section*{Acknowledgments}
The authors are grateful to
V. M. Egu\'iluz for fruitful discussions. 
J.C.G-A acknowledges support from project CNPq 150452/2011-0, Brazil. 
M. G. C. thanks the Associates' Programme
of the Abdus Salam International Center for Theoretical Physics, Trieste, Italy.
M. S. M. acknowledges financial support from Comunitat Aut\`onoma de les Illes
Balears, FEDER, and MINECO (Spain) under project FIS2007-60327


\begin{thebibliography}{99}
\bibitem{Elamar} Elasmar M G, editor (2003) The Impact of International Television: A Paradigm Shift. 
 New Jersey: Lawrence Erlbaum Associates Inc. 213 p.
\bibitem{Bryant} Bryant J, Zillman D ,  editors (2002), Media Effects: Advances in Theory and Research, New Jersey: Lawrence Erlbaum. 648 p.
\bibitem{Nova} Kov\'acs HV, editor (2011), Mass Media: Coverage, Objectivity and Changes, Hauppauge, New York: Nova Science Publishers. 
\bibitem{Yaple} Yaple P, Korzenny F (1989) Electronic mass media effects across cultures. In:  Asante MK, Gudykunst BW, editors.  Handbook of International and Intercultural Communication. Newbury Park, California: Sage.
\bibitem{Chinos} Rhee JW, Lee C (2010) Cross-cultural interactions through mass media products: Cognitive and emotional impacts of Chinese people’s consumption of Korean media products. In: Black D, Epstein S, Tokita A, editors. Complicated Currents: Media Flows, Soft Power and East Asia. Melbourne: Monash University ePress. pp. 5.1-5.16.   
\bibitem{Tan} Tan A, Li S, Simpson C (1986) American television and social stereotypes of Americans in Taiwan and Mexico. Journalism Quarterly 64: 809-814.
\bibitem{Weiman} Weimann G (1984) Images of life in America: The impact of American TV in Israel. International Journal of Intercultural Relations 8: 185-197.
\bibitem{Zaha} Zaharopoulos T (1997) US television and American cultural stereotypes in Greece. World Communication 26: 30-45.
\bibitem{Hunter} Elasmar MG, Hunter JE (2003) A meta-analysis of crossborder effect studies. In:
Elasmar M G, editor. The Impact of International Television: A Paradigm Shift.  New Jersey: Lawrence Erlbaum Associates Inc.  pp, 127-150. 
 \bibitem{Castellano} Castellano C, Fortunato S, Loreto V (2009) Statistical physics of social
dynamics.  Rev. Mod. Phys. 81: 591-646.
\bibitem{Marro} Marro J, Dickman R (1999) Nonequilibrium Phase Transitions in Lattice Models. Cambridge: Cambridge University Press. 327 p. 
\bibitem{Weid} Weidlich W (1991)  Physics and social
science - The approach of synergetics.  Phys. Rep. 204: 1-163. 
\bibitem{Sucheki} Suchecki K, Egu\'iluz VM, San Miguel M (2005) Voter model dynamics in complex networks: Role of dimensionality, disorder, and degree distribution. Phys. Rev. E 72: 036132. 
\bibitem{Stauffer} Stauffer D, de Oliveira SM, de Oliveira PMC, S\'a Martins J (2006) Biology, Sociology, Geology by Computational Physicists. Amsterdam: Elsevier.  
\bibitem{Vilone1} Vilone D, Ramasco J, S\'anchez A; San Miguel M (2012) Social and strategic imitation: the way to consensus. Scientific Reports 2. 
\bibitem{Zimmerman} Zimmerman M, Egu\'iluz VM, San Miguel M (2001) Economics with Heterogeneous Interacting Agents. In: Kirman A, B. Zimmerman BJ, editors. Lecture Notes in Economics and Mathematical Systems 503. Berlin: Springer Verlag, pp. 73-86.
\bibitem{Maxi} Zimmerman M, Egu\'iluz VM, San Miguel M (2004) Coevolution of dynamical states and interactions in dynamic networks. Phys. Rev. E. 69: 065102(R).
\bibitem{Holme} Holme P, Newman MEJ (2006) Nonequilibrium phase transition in the coevolution of networks and opinions. Phys. Rev. E 74: 056108. 
\bibitem{Centola} Centola D, Egu\'iluz VM, Macy WW (2007) Cascade dynamics of complex propagation. Physica A 374: 449-456.
\bibitem{Barto} Bartolozzi M, Leinweber DB, Thomas AW (2005) Stochastic opinion formation in scale-free network. Phys. Rev. E 72: 046113.
\bibitem{Axel} Axelrod R (1997) The dissemination of culture: A model
with local convergence and global polarization. J. Conflict Res. 41: 203-225. 
\bibitem{CMV} Castellano C, Marsili M, Vespignani A (2000) Nonequilibrium phase transition in a model for social influence. Phys. Rev. Lett. 85:  3536-3539.
\bibitem{Vilone} Vilone D, Vespignani A, Castellano C (2002) Ordering phase transition in the one-dimensional Axelrod model.  Eur. Phys. J. B 30: 399-406.
\bibitem{Klemm} Klemm K, Egu\'iluz VM, Toral R, San Miguel M (2003) Nonequilibrium transitions in complex networks: A model of social interaction. Phys. Rev. E 67: 026120.
\bibitem{Klemm2} Klemm K, Egu\'iluz VM,  Toral R, San Miguel M (2003) Global culture: A noise-induced transition in finite systems. Phys. Rev. E 67: 045101(R).
\bibitem{JC1} Gonz\'alez-Avella JC, Cosenza MG, Tucci K (2005) Nonequilibrium transition induced by mass media in a model for social influence. Phys. Rev. E 72: 065102(R). 
\bibitem{JC} Gonz\'alez-Avella JC, Egu\'iluz VM, Cosenza MG, Klemm K, Herrera JL, San Miguel M (2006)
Local versus global interactions in nonequilibrium transitions: A model of social dynamics. Phys. Rev. E 73: 046119. 
\bibitem{Klemm3} Peres LR, Fontanari JF (2011) The media effect in Axelrod's model explained. EPL 96: 38004.
\bibitem{Kuperman} Kuperman MN (2006) Cultural propagation on social networks. Phys. Rev. E 73: 046139. 
\bibitem{Candia} Mazzitello KI, Candia J, Dossetti V (2007) Effects of mass media and cultural drift in a model for social influence. Int. J. Mod. Phys. C 18: 1475-1482. 
\bibitem{Galla} De Sanctis L, Galla T  (2009) Effects of noise and confidence thresholds in nominal and metric Axelrod dynamics of social influence. Phys. Rev. E 79: 046108.
\bibitem{Gracia} Gracia-Lazaro C, Floria LM, Moreno Y (2011) Selective advantage of tolerant cultural traits in the Axelrod-Schelling model. Phys. Rev. E 83: 056103. 
\bibitem{Zhang} Zhang W, Lim C, Sreenivasan S, Xie J, Szymanski BK, Korniss G (2011)  Social influencing and associated random walk models: Asymptotic consensus times on the complete graph. Chaos 21: 025115. 
\bibitem{Mikhailov} Mikhailov AS, Calenbuhr V (2002) From Cells to Societies: Models of Complex Coherent Action.
Berlin: Springer. 299 p.
\bibitem{Deff} Deffuant G, Neau D,  Amblard F, Weisbuch G (2000)  Mixing beliefs among interacting agents. Adv. Complex Syst. 3: 87-98. 
\bibitem{Amblard} Weisbuch G,  Deffuant G,  Amblard F, Nadal JP (2002) Meet, discuss, and segregate.  Complexity 7: 55-63.
\bibitem{Krause} Hegselmann R,  Krause K (2002) Opinion dynamics and bounded confidence: models, analysis and simulation. J. Artif. Soc. Soc. Simul. 5: (3) 2 . http://jasss.soc.surrey.ac.uk/5/3/2.html.
\bibitem{Zanette} Laguna MF, Abramson G, Zanette DH (2003) Vector opinion dynamics in a model for social influence. Physica A 329: 459-472.
\bibitem{Shibanai} Shibanai Y, Yasuno S, Ishiguro I (2001) Effects of Global Information Feedback on Diversity:
Extensions to Axelrod's Adaptative Culture Model. J. Conflict Res. 45: 80-96.
\bibitem{NJP} Gonz\'alez-Avella JC, Cosenza MG,  Egu\'iluz VM, San Miguel M (2010) Spontaneous ordering against an external field in non-equilibrium systems. New J. Phys. 12: 013010. 
\bibitem{Candia2} Candia J, Mazzitello, KI (2008) Mass media influence spreading in social networks with community structure. J. Stat. Mech. Theor. Exp.: P07007.
\bibitem{Arezky1} Rodr\'iguez AH, del Castillo-Mussot M, V\'azquez GJ (2009) Induced monoculture in Axelrod model with clever mass media. Int. J. of Modern Physics C 20: 1233-1245. 
\bibitem{Arezky2} Rodr\'iguez AH, Moreno Y (2010) Effects of mass media action on the Axelrod model with social influence.  Phys. Rev. E 82: 016111.
\bibitem{Gar} Gargiulo F, Lottini S, Mazzoni A (2008) The saturation threshold of public opinion: are aggressive media campaigns always effective?.  arXiv:0807.3937.
\bibitem{Kuramoto} Kuramoto Y, Battogtokh D (2002) Coexistence of coherence and incoherence in nonlocally coupled phase oscillators. Nonlinear Phenom. Complex Syst. 5: 380-385.
\bibitem{Strogatz} Abrams DM, Strogatz SH (2004) Chimera states for coupled oscillators. Phys. Rev. Lett. 93: 174102.
\bibitem{Abrams} Abrams DM, Mirollo R, Strogatz SH, Wiley DA (2008) Solvable model for chimera states of coupled oscillators. Phys. Rev. Lett. 101: 084103. 
\bibitem{Laing} Laing C R  (2010) Chimeras in networks of planar oscillators. Phys. Rev. E 81: 066221
\bibitem{Scholl} Omelchenko I, Maistrenko Y, H\"ovel P, Sch\"oll E (2011) Loss of coherence in dynamical networks: spatial chaos and chimera states. Phys. Rev. Lett. 106: 234102.
\bibitem{Showalter} Tinsley MR, Nkomo S, Showalter K  (2012) Chimera and phase-cluster states in populations
of coupled chemical oscillators. Nature Phys. 8: 662-665.
\bibitem{Roy} Hagerstrom AM, Murphy TE, Roy R,  H\"ovel P,
Omelchenko I,  Sch\"oll E (2012) Experimental observation of chimeras in
coupled-map lattices. Nature Phys. 8: 658-661.
\bibitem{Fede} Vazquez F, Redner S (2007) Non-monotonicity and divergent time scale in Axelrod model dynamics. EPL 78: 18002 
\bibitem{Lima} Mathews CG, Lesku JA, Lima SL, Amlaner CJ (2006) Asynchronous eye closure as an anti-predator behavior in the western fence lizard (Sceloporus Occidentalis). Ethology 112: 286-292. 
\bibitem{Huth}  Huth A, Wissel Ch (1992) The simulation of the movement of
fish shoals.  J. Theor. Biol. 156: 365-385. 
\bibitem{Vicsek} Vicsek T, Czir\'ok A, Ben-Jacob E, Cohen I, Schochet O (1995) Novel type of phase transition in a system of self-driven particles. Phys. Rev. Lett. 75: 1226-1229.
\bibitem{Chate} Gr\'egoire G, Chat\'e H, Tu Y(2003) Moving and staying together without a leader. Physica D 181: 157-170. 
\bibitem{Kaneko} Shibata T, Kaneko K  (2003) Coupled map gas: structure formation and dynamics of interacting motile elements with internal dynamics. Physica D 181: 197-214.
\bibitem{Kenkre} Fuentes MA, Kuperman MN, Kenkre VM (2003) Nonlocal interaction effects on pattern formation in population dynamics. Phys. Rev. Lett. 91: 158104. 
\end{thebibliography}
\end{document}